\documentclass[aps,pre,twocolumn,groupedaddress,showpacs,superscriptaddress]{revtex4}
\usepackage{graphicx}
\usepackage{color}
\usepackage{amsmath}
\usepackage{hyperref}

\begin{document}

\title{Nikola Tesla's Inventions - Myth or Reality?}

\author{Val G.~Rousseau, \url{https://vadcpa.com/Val}}
\affiliation{Department of Cell Biology and Anatomy, LSU Health New Orleans, School of Medicine, New Orleans, Louisiana 70112, USA}

\begin{abstract}
Nikola Tesla is often presented by his adepts as the ``unjustly forgotten genius", without who our current technology wouldn't exist. 
In this paper, we analyze some popular statements made by Tesla's adepts, mostly about inventions that they attribute to him, and determine whether they are myth or reality. \textit{A video of this paper is available\cite{YouTube}}.
\end{abstract}

\pacs{???}
\maketitle

\section{Introduction}
In the last three decades, the development of the Internet has opened to everyone the easy access to a tremendous amount of information. Unfortunately, in parallel, it has allowed people from everywhere to propagate and publish fake information. As a result, it is necessary to take any piece of information found on the web with a grain of salt, and check the sources.

This is especially true when we see the web being flooded with an incredible number of posts and websites praising Nikola Tesla, and presenting him as the genius of all times, making claims about inventions that he supposedly made, without being able to give any single reference to an authentic document or a peer-reviewed publication that could support them.

Nikola Tesla was a Serbian (later naturalized American)\cite{Item1} inventor, electrical engineer, and mechanical engineer. He was also qualified as a ``futurist", and is well known for his contributions to the design of the modern electricity supply system.

Despite the contributions and inventions that he really made, Tesla adopted pseudo-scientific views, as testified by his rejection of various experimental evidences (such as the existence of the electron\cite{Electrons}). In addition, he was stubborn and his refusal for treating physical phenomena with the required mathematical background led him to misunderstand the electromagnetic theory. Indeed, Tesla declared ``a wireless transmitter does not emit Hertz waves which are a myth, but sound waves in the ether"\cite{Electromagnetic}. He also completely missed the advances made in physics in the early 1900s, especially in Relativity and Quantum Mechanics\cite{Electromagnetic,Atomic}. By adopting such a behavior, and publicly making aggressive comments against theories that actually turned out to be correct (especially General Relativity, which yet had already been tested successfully and validated), Tesla destroyed his own fame. This scenario made him the perfect guru for adepts of conspiracy theories who want to believe that it is other scientists who destroyed his fame in order to hide from people the existence of some ``free energy" that Tesla was supposedly trying to offer to the world, although this is in total contradiction with the fact that energy cannot be created nor destroyed.

The following section presents and analyzes some popular statements about Nikola Tesla made by his adepts.

\section{Popular statements by Tesla's adepts}
\subsection{Statement \#1}
\textit{``Thomas Edison stole from Nikola Tesla the invention of the light bulb."}\\

The first electric light was invented by Sir Humphry Davy. He attached a fine charcoal strip between the wires that were connected to a battery. This was in 1809, almost 50 years before Tesla was born\cite{Davy}. In 1840, that is still 16 years before Tesla was born, Warren de la Rue put a platinum filament inside a sealed vacuum tube\cite{DeLaRue}. In 1875, Henry Woodwart and Matthew Evans patented the light bulb\cite{Woodwart}. In 1880, Thomas Edison having purchased Woodwart and Evans' patent, presented a light bulb with a carbonated bamboo filament with a lamp life of 600 hours\cite{Edison}. This was the first real commercial model.

\underline{Verdict}: Myth. Tesla didn't invent the light bulb, and Edison didn't steal the invention. Edison rightfully bought the patent. He also improved the light bulb greatly, so that it became useful and affordable to everyone.\\

\subsection{Statement \#2}
\textit{``Nikola Tesla invented the laser."}\\

The pathway leading to an invention is usually long, and involves the contribution of many actors. In the case of the laser, it started with Albert Einstein in 1917 who settled its foundations\cite{Einstein}, and continued with the work of Rudolf Ladenburg\cite{Ladenburg}, and many others. It is only in 1960 that Theodore H. Maiman operated the first functioning laser\cite{Maiman1,Maiman2}. Einstein's laser effect is a prediction based on Quantum Mechanics, which Tesla strongly rejected.

\underline{Verdict}: Myth. The laser is inherently a quantum mechanical device, and there is no way that Tesla could have invented it with his erroneous views of classical physics.

\subsection{Statement \#3}
\textit{``Nikola Tesla invented the electric motor."}\\

The electric motor was invented by Michael Faraday in 1821, 35 years before Tesla was born\cite{Faraday1}.

\underline{Verdict}: Myth. The electric motor was clearly invented before Tesla was born.

\subsection{Statement \#4}
\textit{``Nikola Tesla invented the AC current."}\\

This one is really well anchored! The ideas behind AC (alternating current) were well established before Tesla was born. In 1832, Michael Faraday demonstrated that three types of electricity that were thought to be different (electricity induced by a magnet in a coil, electricity produced by a battery, and static electricity) were in fact all the same\cite{Faraday2}. In particular, the direction of the current induced in a coil by a magnet depends on whether the magnet is approached or moved away. Thus, Faraday knew that, by moving the magnet back and forth, an alternating current was produced. The first alternator to produce alternating current was a dynamo electric generator based on Michael Faraday's principle, constructed by French instrument maker Hyppolite Pixii the same year\cite{Pixii}.

\underline{Verdict}: Myth. Tesla clearly did not invent AC current. The first alternator was built 24 years before he was born.

\subsection{Statement \#5}
\textit{``Nikola Tesla invented the AC motor."}\\

The first AC induction motor was invented in 1885 by Galileo Ferraris\cite{Ferraris}. Ferraris demonstrated a working model of his invention before Tesla came up with a modified version.

\underline{Verdict}: Myth. Tesla's motor was certainly different, and deserves some praise, but it came after Ferraris's.

\subsection{Statement \#6}
\textit{``Nikola Tesla invented the wireless communication."}\\

Electromagnetic waves were predictedin 1865 by James Clerk Maxwell's equations\cite{Maxwell}. The existence of Electromagnetic waves was purely theoretical, until Heinrich Hertz in 1886 proved that Maxwell's prediction was correct\cite{Hertz}. For this, he applied a high voltage to a coil, and noticed that a second coil, several feet away, was generating a spark. Unfortunately, Hertz didn't see the potential of his discovery. In 1894, Guglielmo Marconi made a demonstration of a radio transmitter and receiver that made a bell ring by pressing a button with no physical connection to the receiver\cite{Marconi}. This was the first radio-controlled device. In 1898, Tesla made a demonstration of his famous radio-controlled boat at Madison Square Garden in New York\cite{Tesla}.

\underline{Verdict}: Myth. Tesla's radio-controlled boat certainly constitutes some achievement, but it came 4 years after Marconi's radio-controlled bell, thus wireless communication was definitely not Tesla's invention.

\subsection{Statement \#7}
\textit{``Nikola Tesla invented the transformer."}\\

In 1831, Michael Faraday performed experiments on induction between coils of wire, including winding a pair of coils around an iron ring, thus creating the first toroidal closed-core transformer\cite{Faraday2}. He is not credited for the invention of the transformer because he only applied pulses of current to the device instead of AC current, and missed the principle that the voltage generated is proportional to the number of turns. The first transformer was actually developed in 1878 by the Ganz company in Budapest\cite{Ganz}.

\underline{Verdict}: Myth. When the first transformer was developed by the Ganz company, Tesla was still in school and hadn't even begun his first job at a telephony company.

\subsection{Statement \#8}
\textit{``Einstein was once asked what it was like to be the smartest man alive. He replied `I don't know, you'll have to ask Nikola Tesla.'"}\\

This statement is very well anchored, however, there is absolutely no evidence that such a conversation between Einstein and his interviewer ever occurred. Nevertheless, it is worth to mention that Tesla vigorously and publicly criticized Einstein and continuously attempted to discredit him by denouncing his work and his theories. He announced publicly\cite{Electromagnetic}:\\

\textit{“Einstein's relativity work is a magnificent mathematical garb which fascinates, dazzles and makes people blind to the underlying errors. The theory is like a beggar clothed in purple whom ignorant people take for a king.”}\\

Einstein refrained from making any comments or rebuttals. He was confident in his own work, and he was not known for openly criticizing anyone, but should he have a chance for a tongue-in-cheek, it would be totally understandable that he could have given the slightly facetious answer ``you'll have to ask Nikola Tesla", with the end of the answer in his mind, ``because the arrogant prick believes he's the smartest man alive".

\underline{Verdict}: Probably a myth, and should it be true, it would have to be understood as sarcasm.

\section{Conclusion}
Nikola Tesla wasn't the genius that adepts of conspiracy theories try to make us believe. Because of his stubbornness, Tesla adopted pseudo-scientific views, was arrogant and aggressive towards other people's work, and completely missed the advances made in physics in the early 1900s, such as in Relativity and Quantum Mechanics\cite{Electromagnetic,Atomic}. He certainly made technological contributions and inventions, but the path leading to an invention usually involves the contributions from numerous predecessors who deserve to be credited for their work.

\begin{acknowledgments}
We would like to thank U.~Sam for financial support.
\end{acknowledgments}


\begin{thebibliography}{1}
	\bibitem{YouTube} V.G. Rousseau -- ``Nikola Tesla - Myth or Reality": \url{https://youtu.be/vLyBzj8CHgM}
	\bibitem{Item1} Michael Burnan, ``Nikola Tesla: Physicist, Inventor, Electrical Engineer", Capstone 2009, ISBN: 0756540860, 9780756540869.
	\bibitem{Electrons} Popular Science Monthly, November 1928, page 171.
	\bibitem{Electromagnetic} New York Herald Tribune, 9/11/1932.
	\bibitem{Atomic} The New York Times, (Section 2, 1), July 2, 1931.
	\bibitem{Davy} William Slingo and Arthur Brooker, Electrical Engineering for Electric Light Artisans. London: Longmans, Green and Co. p. 607.
	\bibitem{DeLaRue} Marc Tyler. The Light Bulb. Mankato Minn.: Capstone, 2004.
	\bibitem{Woodwart} Library and Archives Canada. "Electric Light." Patent no. 3738, filed by Henry Woodward and Mathew Evans, 1874.
	\bibitem{Edison} Thomas Edison's Patent Application for the Light Bulb: \url{https://www.archives.gov/historical-docs/edisons-light-bulb-patent}
	\bibitem{Einstein} Einstein, A (1917). ``Zur Quantentheorie der Strahlung". Physikalische Zeitschrift. 18: 121–128.
	\bibitem{Ladenburg} Steen, W.M. "Laser Materials Processing", 2nd Ed. 1998.
	\bibitem{Maiman1} Townes, Charles Hard. "The first laser". University of Chicago. Retrieved May 15, 2008.
	\bibitem{Maiman2} Hecht, Jeff (2005). Beam: The Race to Make the Laser. Oxford University Press. ISBN 978-0-19-514210-5.
	\bibitem{Faraday1} Nguyen, Tuan C. "A Biography of Michael Faraday, Inventor of the Electric Motor." ThoughtCo, Oct. 28, 2020.
	\bibitem{Faraday2} Michael Faraday, ``Experimental Researches in Electricity", ISBN-13: 978-0486435053.
	\bibitem{Pixii} Mohamed A. El-Sharkawi, Electric Energy: An Introduction, Third Edition, CRC Press, 2015, ISBN 1498760031,page 3.
	\bibitem{Ferraris} Dufresne, Steven, ``Inventing The Induction Motor" (September 21, 2017).
	\bibitem{Maxwell} Maxwell, James Clerk, ``A dynamical theory of the electromagnetic field" (1865). Philosophical Transactions of the Royal Society of London. 155: 459–512. 
	\bibitem{Hertz} Baird, Davis, Hughes, R.I.G. and Nordmann, Alfred eds. (1998). Heinrich Hertz: Classical Physicist, Modern Philosopher. New York: Springer-Verlag. ISBN 0-7923-4653-X. p. 53.
	\bibitem{Marconi} Brown, Antony. Great Ideas in Communications. D. White Co., 1969, page 141.
	\bibitem{Tesla} Nikola Tesla, Madison Square Garden, New York 1998 (U.S. Patent 613,809).
	\bibitem{Ganz} Guarnieri 2013, pp. 56–59, Hughes 1993, pp. 95–96.
\end{thebibliography}
\end{document}